\documentclass[aps,prd,amsmath,amssymb,preprintnumbers,preprint,nofootinbib,a4paper,11pt]{revtex4-1}
\pdfoutput=1
\usepackage{graphicx}
\usepackage{url}
\usepackage[bookmarks, pagebackref=true]{hyperref}
\usepackage{color}
	\definecolor{rossoCP3}{cmyk}{0,.88,.77,.40}
		\definecolor{graa}{rgb}{0.8,0.8,0.8}
		\definecolor{blaa}{rgb}{0.2,0.2,0.6}
		\hypersetup{
			colorlinks, 
			bookmarksopen, 
			bookmarksnumbered,
			citecolor=blaa, 		
			linkcolor=rossoCP3,	
			urlcolor=rossoCP3,			
			}
\usepackage{bm}
\usepackage{bbm}
\usepackage{pxfonts}
\usepackage[margin=5pt, font=small,labelfont=bf,justification=raggedright]{caption}
\usepackage{subfig}
\usepackage{youngtab}
\usepackage{slashed}

\usepackage{feynmf}           

\newcommand{\beq}{\begin{eqnarray}}
\newcommand{\eeq}{\end{eqnarray}}

\newcommand{\bmp}{\noindent\begin{minipage}{16cm}}
\newcommand{\emp}{\end{minipage}\vskip 7mm} 

\usepackage{mathrsfs}         

\begin{document}
\phantom{g}\vspace{2mm}
\title{ \Large  \color{rossoCP3}  Dark Matter Interference} 
  \author{Eugenio  {\sc Del Nobile}}
  \email{delnobile@cp3-origins.net}
  \author{Chris  {\sc Kouvaris}}
  \email{kouvaris@cp3-origins.net}
  \author{Francesco {\sc Sannino}}
  \email{sannino@cp3-origins.net}
  \author{Jussi  {\sc Virkaj\"{a}rvi}} 
  \email{virkajarvi@cp3-origins.net}
\affiliation{
{\large \rm \color{rossoCP3}CP$^3$-Origins} 
\&  
{\large D}anish {\large I}nstitute {\large  f}or {\large  A}dvanced {\large  S}tudy {\color{rossoCP3}\large \rm DIAS},
\\
University of Southern Denmark, Campusvej 55, DK-5230 Odense M, Denmark}
\begin{abstract}
 We study different patterns of interference in WIMP-nuclei elastic scattering that can accommodate the  DAMA and CoGeNT experiments via an isospin violating ratio $f_n/f_p=-0.71$.  We study interference between the following pairs of mediators: $Z$ and $Z'$,  $Z'$ and Higgs, and two Higgs fields. We show under what conditions interference works. We also demonstrate that in the case of the two Higgs interference, an explanation of the DAMA/CoGeNT is consistent with Electroweak Baryogenesis scenarios based on two Higgs doublet models proposed in the past.
 \\[.1cm]
{\footnotesize  \it Preprint: CP$^3$-Origins-2011-39 \& DIAS-2011-32.}
\end{abstract}

\maketitle

\section{Introduction}

Asymmetric dark matter has emerged as a competitive paradigm to thermally produced dark matter. Instead of having a mechanism where the population of Weakly Interacting Massive Particles (WIMPs) is controlled by annihilations, it is possible to have an initial asymmetry between particles and antiparticles. The existence of a conserved quantum number associated with the WIMPs can protect them from decay or co-annihilation. If the particle-antiparticle annihilation is sufficiently strong, antiparticles are eliminated by equal number of particles, and due to the asymmetry, the dark matter (DM) consists of the remaining particles. If annihilations are not strong enough, a mixed case with a substantial number of antiparticles present today is also possible~\cite{Belyaev:2010kp}. Obviously in this mixed scenario DM consists of both particles and antiparticles. The fact that the asymmetry in the DM sector resembles the baryonic asymmetry, makes asymmetric DM easily incorporated in extensions of the Standard Model. It also means that the asymmetries in the baryonic and dark sector might be related. Although the idea of asymmetric DM is not new
\cite{Nussinov:1985xr,Barr:1990ca,Gudnason:2006ug,Gudnason:2006yj}, it has recently attracted a lot of  interest~\cite{ Foadi:2008qv,Khlopov:2008ty,Khlopov:2008ty,Dietrich:2006cm,Sannino:2009za,Ryttov:2008xe,Sannino:2008nv,Kaplan:2009ag,Frandsen:2009mi,MarchRussell:2011fi,Frandsen:2011kt,Gao:2011ka,Arina:2011cu,Buckley:2011ye,Lewis:2011zb,Davoudiasl:2011fj,Graesser:2011wi,Bell:2011tn}.

\begin{figure}[h!]
\begin{center}
\includegraphics[width=.49\textwidth
]{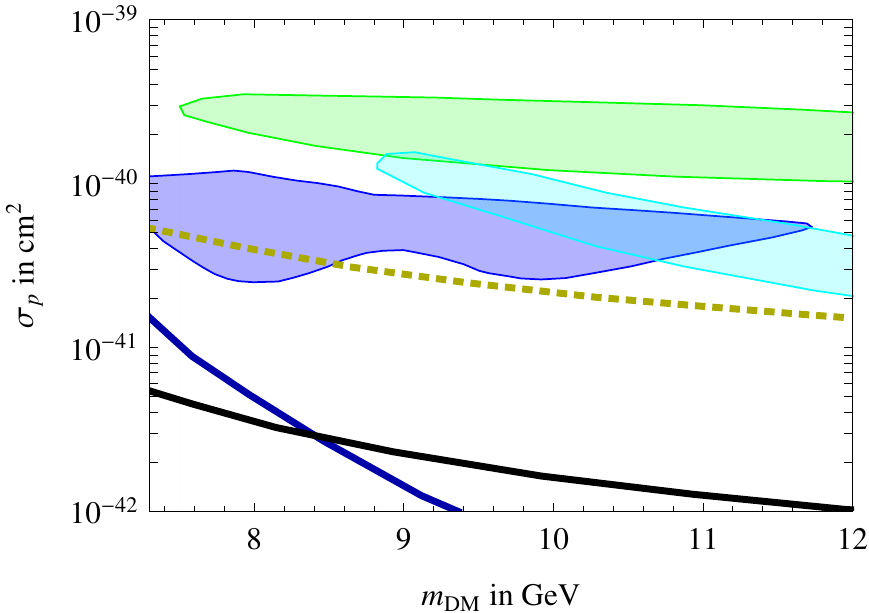}
\includegraphics[width=.49\textwidth
]{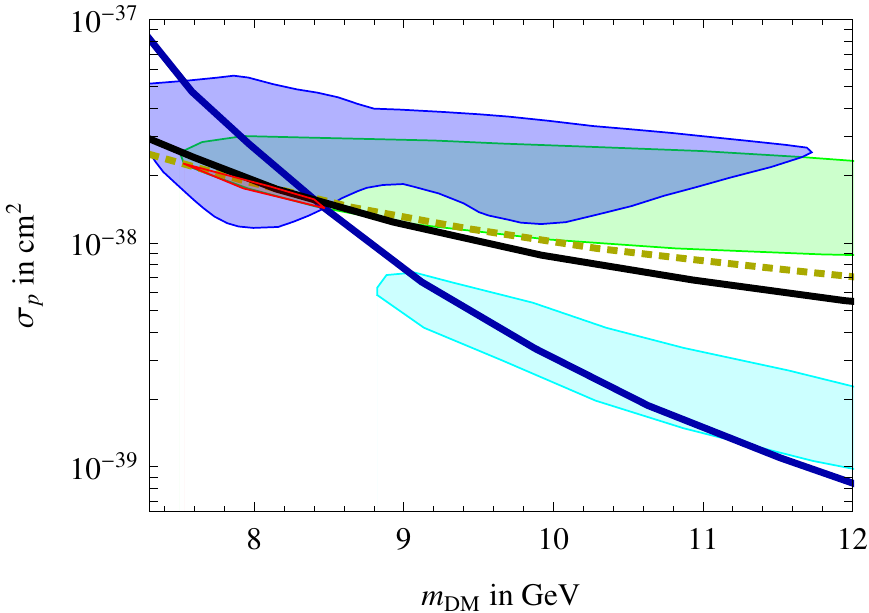}
\caption{\em\label{fig:experiments}Favored regions and exclusion contours in the $(m_{DM}, \sigma_p)$ plane for the standard case $f_n / f_p = 1$ (left panel) and the case $f_n / f_p = - 0.71$ (right panel). The green contour is the $3 \sigma$ favored region by DAMA \cite{Savage:2010tg} assuming no channeling \cite{Bozorgnia:2010xy} and that the signal arises entirely from Na scattering; the blue region is the $90 \%$ CL favored region by CoGeNT; the cyan contour is the $2 \sigma$ favored region by CRESST-II~\cite{Angloher:2011uu}; the dashed line is the exclusion plot by CDMS II Soudan \cite{Ahmed:2010wy}; and the black and blue lines are respectively the exclusion plots from the Xenon10 \cite{Angle:2011th} and Xenon100 \cite{Aprile:2011hi} experiments. The CoGeNT and DAMA overlapping region passing the constraints is shown in red.}
\end{center}
\end{figure}

The current status of experimental direct detection of DM is quite intriguing. A signal with annual modulation possibly attributed to DM has been solidly established in  DAMA~\cite{Bernabei:2008yi}, and more recently in CoGeNT~\cite{Aalseth:2011wp}. In addition CRESST-II~\cite{Angloher:2011uu} has recently released results compatible with the existence of a light WIMP too. However, experiments such as CDMS ~\cite{Ahmed:2010wy}, and Xenon10/100~\cite{Angle:2011th,Aprile:2011hi} find null evidence for DM, imposing thus severe constraints on WIMP-nucleons cross sections. The fact that some experiments detect DM and some other do not is not the only experimental discrepancy one faces. Upon assuming spin-independent interactions between WIMPs and nuclei, it is clear that DAMA and CoGeNT are at odds, if WIMPs couple to protons and neutrons with equal strength.  However if the relative couplings of WIMPs to neutrons and protons satisfy $f_n/f_p\simeq -0.71$ ~\cite{Chang:2010yk,Feng:2011vu}, an agreement of DAMA and CoGeNT is possible, and it indicates a DM-proton cross section $\sigma_p \sim 2 \times 10^{-38} \rm{ cm}^2$ and a DM mass $m_{DM} \approx 8$ GeV (see Fig.~\ref{fig:experiments}). In a recent paper~\cite{DelNobile:2011je} we demonstrated that the isospin violation 
needed to produce $f_n/f_p\simeq -0.71$ can be easily accommodated using Standard Model mediators, via interference of two different channels in elastic WIMP-nuclei collisions (see also \cite{Cline:2011zr,Cline:2011uu}). Interfering DM can thus naturally explain the above phenomenological ratio. We showed that if Interfering DM is made of composite asymmetric WIMPs (with electroweak compositeness scale), a simple interference in the WIMP-nucleus collision between a photon exchange (via a dipole type interaction) and a Higgs exchange can produce the required isospin violation. Interestingly this candidate has been proven to arise in strongly interacting models using first principle lattice computations~\cite{Lewis:2011zb}.  

In this paper we extend the idea of interfering DM by presenting three general interference patterns for fermionic DM that can accommodate the experimental findings. More specifically we show under what conditions  interference between $Z$ and a $Z'$; $Z'$ and Higgs, and two Higgs doublets can provide the appropriate isospin violation. In the last case we show that the interference between the two Higgs scalars can also be compatible with Electroweak Baryogenesis~\cite{Cohen:1991iu,Nelson:1991ab,Joyce:1994zn}.  
We should also mention that  observations of neutron stars put severe constraints on the spin-dependent cross section of fermionic asymmetric WIMPs~\cite{Kouvaris:2010jy}, and bosonic asymmetric WIMPs~\cite{Kouvaris:2011fi}. In our study here we avoid these constraints because our fermionic asymmetric WIMP candidates do not have significant spin-dependent cross section.


\section{$Z$ interfering with $Z'$}

First we will consider a scenario where a fermionic DM particle $\psi$ couples to the $Z$-boson and to a spin-1 state $Z'$. 

The $Z$-DM and $Z$-nucleon interaction Lagrangian, including only renormalizable terms, reads 
\begin{align}\label{Zint}
\mathscr{L}_Z = \,&\frac{g}{2 \cos \theta_W} Z_{\mu} \bar\psi (v_{\psi}
 -  a_{\psi} \gamma^5) \gamma^\mu\psi \,  + \\ \nonumber 
&\frac{g}{2 \cos \theta_W} Z_{\mu} \left[ \bar p\, \gamma^\mu (v_p - a_p \gamma^5)  p \, + \bar n\, \gamma^\mu (v_n - a_n  \gamma^5) n \right] \ ,
\end{align}
where the $Z$-DM couplings $v_{\psi}$ (vector) and $a_{\psi}$ (axial-vector) are normalized to the usual weak coupling strength. $p$ and $n$ refer respectively to protons and neutrons and the $Z$-nucleon vector and axial-vector couplings are  
\begin{align}
 v_p =  \frac{1}{2} - 2\sin^2 \theta_W \,,\quad  v_n =  -\frac{1}{2} \,,\quad a_p  = 1.36 \,,\quad a_n  = -1.18 \,, \nonumber 
\end{align}
where we have used the numerical values from  \cite{Ellis:2008hf} to estimate $a_p$ and $a_n$. 
However, we are not concerned with the axial-vector couplings, since their contribution to the cross section is suppressed with respect to the one given by the vector couplings.
Similarly the $Z'$-DM and $Z'$-nucleon interaction Lagrangian can be written as
\begin{align}\label{Zpint}
\mathscr{L}_{Z'} = \, &\frac{g}{2 \cos \theta_W} Z'_{\mu} \bar\psi (v'_{\psi} - a'_{\psi} \gamma^5) \gamma^\mu\psi \,+ \\ \nonumber 
& \frac{g}{2 \cos \theta_W} Z'_{\mu} 	\left[ \bar p\, \gamma^\mu (v'_p - a'_p \gamma^5) \,p \, + \bar n\, \gamma^\mu (v'_n - a'_n  \gamma^5)\ n \right] \ .
\end{align}
As for the $Z$, also in this case the axial-vector couplings contribution to the cross section is negligible.
Possible constraints from colliders on $Z'$ can be safely avoided assuming a leptophobic $Z'$. 
As long as the $Z'$ couplings to leptons are small enough, no bounds can be set at present. Using  Eqs.~\eqref{Zint} and \eqref{Zpint}, we can write the spin-independent cross section in the zero momentum transfer limit as
\begin{align}\label{sigmaZZp} 
\sigma = \frac{ 2 G_F^2\mu_{A}^2}{\pi}  \left| f_p \, \mathsf  Z + f_n (\mathsf A-\mathsf Z)  \right|^2,
\end{align}
where  $G_F$ is the Fermi constant, $\mu_A$ is the DM-nucleus reduced mass, and the dimensionless couplings to protons and neutrons are defined as
\beq
 f_p =  v_{\psi} v_p +  v'_{\psi}v'_p \frac{m^2_{Z}}{m^2_{Z'}}   \,,\qquad
 f_n =  v_{\psi} v_n   +v'_{\psi}v'_n  \frac{m^2_{Z}}{m^2_{Z'}}    \,.
 \eeq
We already know that in order to alleviate the discrepancy between the different direct detection experimental results we need to 
have $m_{DM} \sim 8$ GeV, $f_n / f_p=-0.71$, and the DM-proton cross section $\sigma_p \sim 2 \times 10^{-38}$ cm$^2$. 
Thus by fixing these three values we find the following relations for the unknown parameters of the model
\begin{align}
 \left| f_p \right| &=  \left|v_{\psi} v_p + v'_{\psi}v'_p \frac{m^2_{Z}}{m^2_{Z'}}\right|  = \sqrt{\frac{ \sigma_p \pi}{2 G_F^2\mu_{p}^2}}  \ ,\\
   f_n &=  v_{\psi} v_n   +  v'_{\psi}v'_n  \frac{m^2_{Z}}{m^2_{Z'} }= - 0.71f_p \ .
 \end{align}
Substituting the numbers and dividing by the known values of the parameters  $v_p = 0.055$ and $v_n = -0.5$ we get the two constraints  
\begin{align}\label{constZZp}
v_{\psi} +  v'_{\psi}  \frac{v'_p}{v_p}\frac{m^2_{Z}}{m^2_{Z'}} &= v_{\psi} + 15 \,v'_{\psi} v'_p  \left(\frac{m_{Z'}}{100\, \rm{GeV}} \right)^{-2} = \pm 17\,, \\
    v_{\psi} +  v'_{\psi} \frac{v'_n}{v_n} \frac{m^2_{Z}}{m^2_{Z'}}  &= v_{\psi} -1.7 \,v'_{\psi}v'_n \left(\frac{m_{Z'}}{100\, \rm{GeV}} \right)^{-2} =  \pm 1.3 \ . \label{fnZZp}
\end{align}
The $Z$-DM coupling $v_{\psi}$ can be constrained using the measurements of the $Z$ decay width into invisible channels. The LEP experiment set strict limits on the number of SM neutrinos, i.e.~$N_{\nu}=2.984\pm0.008$ \cite{Nakamura:2010zzi}.  The error in the measurement can be used to constrain non-SM contributions to the $Z$ decay width. Using the uncertainty in the LEP result $\delta_{\rm LEP}=0.008$, this yields
\beq
  v_{\psi}^2 \beta (3-\beta^2) + 2 a_{\psi}^2 \beta^3 < \delta_{\rm LEP} \, ,
\eeq 
where $\beta = \sqrt{1-4 m^2_{DM}/m^2_Z}$ is the velocity factor. 
Assuming a DM mass of $\sim 8$ GeV, with no axial-vector coupling ($a_{\psi} = 0$), the vector coupling $v_{\psi}$ can assume its maximal allowed value $|v_{\psi}| < 0.063$, while for $a_{\psi} = v_{\psi}$ this constraint gives $|v_{\psi}| < 0.045$. Taking into account this strong bound in  Eqs.~\eqref{constZZp}, and~\eqref{fnZZp}, it is evident that the bulk contribution is due to the $Z'$ alone. Therefore  interference is not relevant for this kind of DM interaction with the SM particles. Similar studies have been performed recently in \cite{Frandsen:2011cg}.

\section{$Z'$ interfering with Higgs}

Before proceeding, let us comment, that the DM signals seen in DAMA/CoGeNT and the null results of the other direct DM experiments cannot be explained simultaneously through a $Z$ and Higgs interference. The reason for this is that a light ($\sim 8$ GeV) Dirac DM particle, with a coupling to the $Z$-boson such that $\sigma_p \sim 2 \times 10^{-38}$ cm$^2$ and $f_n/f_p = -0.71$, is ruled out by the aforementioned LEP constraints. However, as we will demonstrate below, interference between $Z'$ and the Higgs is a viable possibility. 


The relevant Higgs ($h$) interaction Lagrangian is
\begin{align}
\mathscr{L}_h = m_{DM}\bar{\psi} \psi -\,  h \bar{\psi} (d_h+a_h \gamma^5)\psi 
-\frac{m_p}{v_{EW}} f \, h \,( \bar p p +\bar n  n ) \  ,
\end{align}
where  $d_{h}$ and $a_{h}$ are the dimensionless scalar and pseudo-scalar Higgs-DM couplings respectively. The Higgs field $h$ is here the physical field, i.e.~the oscillation around the vacuum expectation value $v_{EW}$. We have specified a mass term for the DM to point out that it doesn't need to be generated by the vacuum expectation value of the Higgs field.

Combining the scalar interaction from this Lagrangian, with the vector one for the $Z'$ as it appears in \eqref{Zpint}, we get the DM-nucleus spin-independent cross section as in Eq.~\eqref{sigmaZZp}, 
where now the dimensionless couplings to protons and neutrons are defined as
\beq
 f_p =   v'_{\psi}v'_p \frac{m^2_{Z}}{m^2_{Z'}} - d_h \frac{f m_p v_{EW}}{  m^2_h}   \,,\qquad
 f_n =    v'_{\psi}v'_n\frac{m^2_{Z}}{m^2_{Z'}} - d_h \frac{f m_p v_{EW}}{  m^2_h}   \,.
 \eeq
 As for the $Z$-$Z'$ case, the pseudo-scalar and pseudo-vector couplings of the DM with the Higgs and the $Z'$ respectively lead to negligible contributions to the cross section compared to the scalar and vector ones investigated here.
The constraints are
\begin{align}
 |f_p| &=  \left| v'_{\psi}v'_p \frac{m^2_{Z}}{m^2_{Z'}} - d_h \frac{f m_p v_{EW}}{  m^2_h} \right|  = 0.92 \\
   f_n &=   v'_{\psi}v'_n\frac{m^2_{Z}}{m^2_{Z'}} - d_h \frac{f m_p v_{EW}}{  m^2_h}  = - 0.71f_p  = \pm0.65 \,.
\end{align}
 These can be rewritten as  
 \begin{align}
 & v'_{\psi}v'_p \left(\frac{m_{Z'}}{100\, \rm{GeV}} \right)^{-2}  - 8.3 \times 10^{-3} d_h \left(\frac{m_h}{100\, \rm{GeV}} \right)^{-2}   = \pm1.1 \\
 & v'_{\psi}v'_n \left(\frac{m_{Z'}}{100\, \rm{GeV}} \right)^{-2}  - 8.3 \times 10^{-3} d_h \left(\frac{m_h}{100\, \rm{GeV}} \right)^{-2}   = \mp0.78 \ ,
\end{align}
where we have used $f = 0.3$~ \cite{Shifman:1978zn}.
\\
If all the couplings are of order unity and $m_{Z'}, m_h \sim 100$ GeV, the Higgs contribution to the interference is negligible, and the $Z'$ has to directly account for the isospin violation needed to get the desired value of $f_n / f_p$. A substantially lighter Higgs, around $50$~GeV with a coupling $d_h$ in the range $5 - 10$, can lead to a phenomenologically viable interference. Note that such a light Higgs-like state is not immediately ruled out by collider experiments since this state has new decay modes, e.g.~to two DM particles which are not accounted for in the SM (see e.g.~\cite{Kim:2008pp}).

\section{Interference within the Two Higgs doublet model} \label{sec2higgs}

We will now consider a two Higgs doublet model where one of the Higgs fields couples to up-type quarks and the other to down-type quarks. This kind of scenario albeit more general, is similar to the Minimal Supersymmetric Standard Model Higgs sector. We consider Yukawa-type interactions between the two Higgs fields and the fermionic DM $\psi$.  We write also effective interactions with the SM proton $p$ and neutron $n$. The interaction Lagrangian is
\begin{align}\label{twohiggs}
\mathscr{L}_{2H}  = \, \lambda^{DM}_1 h_1 \bar\psi \psi \,+ \lambda^{p}_1  h_1 \bar p  p \, +\, \lambda^{n}_1  h_1 \bar n n    
  + \,\lambda^{DM}_2  h_2 \bar\psi \psi \,  +
 \lambda^{p}_2  h_2 \bar p  p \, + \,  \lambda^{n}_2 h_2 \bar n n \ .
\end{align}
$ h_1$ and $ h_2$ are here the physical scalars, i.e.~the mass eigenstates after diagonalization, where the original Higgs fields coupled one to the up-type quarks and the other to the down-type. The nucleon couplings are then
\begin{subequations}\label{couplings1}
\begin{align}
\lambda^{p}_1 &= \frac{ \cos\theta}{v_1} \sum_{q_u} \langle p |m_{q_u} \bar{q}_u q_u |p \rangle -  \frac{\sin\theta}{v_2} \sum_{q_d}  \langle p |m_{q_d} \bar{q}_d q_d |p \rangle \ ,
\\
\lambda^{n}_1 &=  \frac{\cos\theta}{v_1} \sum_{q_u} \langle n |m_{q_u} \bar{q}_u q_u |n \rangle -  \frac{\sin\theta}{v_2} \sum_{q_d}  \langle n |m_{q_d} \bar{q}_d q_d |n \rangle \ ,
\\
\lambda^{p}_2 &=\frac{ \sin\theta }{v_1} \sum_{q_u} \langle p |m_{q_u} \bar{q}_u q_u |p \rangle +  \frac{\cos\theta}{v_2} \sum_{q_d}  \langle p |m_{q_d} \bar{q}_d q_d |p \rangle \ ,
\\
\lambda^{n}_2 &=  \frac{\sin\theta}{v_1} \sum_{q_u}  \langle n |m_{q_u} \bar{q}_u q_u |n \rangle + \frac{ \cos\theta}{v_2} \sum_{q_d}  \langle n |m_{q_d} \bar{q}_d q_d |n \rangle \ ,
\end{align}
\end{subequations}
where the sums over up-type ($q_{u}$) and down-type ($q_{d}$)  quarks account for the scalar quark currents within the nucleons. $v_1$ and $v_2$ are the vacuum expectation values of the two Higgs fields, which obey the relation $v^2_{EW}/2=v^2_1+v^2_2$ ($v_{EW}\simeq 246$~GeV). $\theta$ is the mixing angle needed to diagonalize the Higgs system, and here is a free parameter. We also assume that the DM particle mass is not generated by the vacuum expectation values of the Higgs fields.  
The matrix elements $ \langle p,n |m_{q_{u,d}} \bar{q}_{u,d} q_{u,d} |p,n \rangle$ in~\eqref{couplings1} are obtained in chiral perturbation theory, when dealing with light quarks, using the measurements of the pion-nucleon sigma term~\cite{Cheng:1988im}, and in the case of heavy quarks, from the mass of the nucleon via trace anomaly~\cite{Shifman:1978zn, Vainshtein:1980ea}.
The experimental uncertainties, especially in the pion-nucleon sigma term, give rise to differences in the values of these matrix elements. As long as  $\lambda^p_i$ and $\lambda^n_i$ are not identical, isospin violation can always be guaranteed.
To evaluate the matrix elements we follow Ref.~\cite{Ellis:2008hf} which makes use of the results found in~\cite{Shifman:1978zn, Vainshtein:1980ea, Cheng:1988im}.  
\begin{subequations}\label{couplings}
\begin{align}
\sum_{q_u} \langle p |m_{q_u} \bar{q}_u q_u |p \rangle \approx 105 \, {\rm MeV} \ ,
&\qquad\qquad
\sum_{q_d}  \langle p |m_{q_d} \bar{q}_d q_d |p \rangle \approx 417 \, {\rm MeV} \ ,
\\
\sum_{q_u} \langle n |m_{q_u} \bar{q}_u q_u |n \rangle \approx 100 \, {\rm MeV} \ ,
&\qquad\qquad
\sum_{q_d}  \langle n |m_{q_d} \bar{q}_d q_d |n \rangle \approx 426 \, {\rm MeV} \ .
\end{align}
\end{subequations}

The spin-independent DM-nucleus cross section can now be calculated using the interaction terms from Eq.~\eqref{twohiggs} 
\begin{align}\label{sigma2H}
\sigma = \frac{\mu_{A}^2}{\pi}  \left| f_p \, \mathsf  Z + f_n (\mathsf A-\mathsf Z)  \right|^2,
\end{align}
where the couplings to protons and neutrons are\footnote{The normalization for $f_p$ and $f_n$ here is different than the one used in Eq.~\eqref{sigmaZZp}.}
\beq \label{fnfp}
 f_p =   \frac{\lambda^{DM}_1 \lambda^{p}_1}{m^2_{ h_1 }} +\frac{\lambda^{DM}_2 \lambda^{p}_2}{m^2_{ h_2 }}    \,,\qquad
 f_n = \frac{\lambda^{DM}_1 \lambda^{n}_1}{m^2_{ h_1 }}   +  \frac{\lambda^{DM}_2 \lambda^{n}_2}{m^2_{ h_2 }}    \,.
 \eeq
Eqs.~\eqref{sigma2H} and~\eqref{fnfp} can be used to study the effects of the interference in a generic two Higgs doublet model.
Substituting the couplings from~\eqref{couplings1}, \eqref{couplings} into~\eqref{fnfp}, and imposing the fitting values for $m_{DM} $, $f_n / f_p$ and  $\sigma_p$, we get
the following constraint equations for the unknown parameters:
\begin{align}\label{2Hconst1}
   &\lambda^{DM}_1 \left(\frac{v_1}{v_{EW}} \right)^{-1}
   \left(   \cos\theta - 4.0\, \sin\theta \frac{v_1}{v_2}\right)
    \left(\frac{m_{h_1}}{100\, \rm{GeV}} \right)^{-2} + \nonumber \\
   & 4.0 \,\lambda^{DM}_2   \left(\frac{v_2}{v_{EW}} \right)^{-1}  
    \left(   \cos\theta + 0.25\, \sin\theta \frac{v_2}{v_1}\right)
     \left(\frac{m_{h_2}}{100\, \rm{GeV}} \right)^{-2}
   = \pm 3.5 \times 10^2 \ ,
   \\
   \rule{0cm}{1cm}
   \label{2Hconst2}
  &\lambda^{DM}_1 \left(\frac{v_1}{v_{EW}} \right)^{-1}
  \left(   \cos\theta - 4.3 \sin\theta \frac{v_1}{v_2}\right)
    \left(\frac{m_{h_1}}{100\, \rm{GeV}} \right)^{-2}  + \nonumber \\
  & 4.3 \,\lambda^{DM}_2   \left(\frac{v_2}{v_{EW}} \right)^{-1}
    \left(   \cos\theta + 0.23 \sin\theta \frac{v_2}{v_1}\right)
     \left(\frac{m_{h_2}}{100\, \rm{GeV}} \right)^{-2}
   = \mp 2.6 \times 10^2  \ .
\end{align}
For natural values of $v_1$ and $v_2 $, i.e.~of the order of $v_{EW}$, and for $ m_{h_1}$ and $m_{h_2}$ of the order of 100-1000 GeV, the DM couplings to the Higgs fields need to be of $ \mathscr{O}(10^3)$ to fit the data.  This large couplings are of course unnatural as such. Thus the original DM-Higgs interactions and related couplings, introduced in Eq.~\eqref{twohiggs}, need to be considered as a simple effective description.  

We will introduce now a model that will accommodate such large values for the effective couplings, and link it also to Electroweak Baryogenesis  \cite{Cohen:1991iu, Nelson:1991ab, Turok:1990in, McLerran:1990zh, Dine:1990fj}. We start by recalling the three Sakharov conditions needed for successful production of a baryon asymmetry 
for the model considered here~\cite{Cohen:1991iu}:
the baryon number violation originates from SM sphalerons; out-of-equilibrium conditions are generated by bubble nucleation in a strong first order electroweak (EW) phase transition;  a new CP violating phase which we take it to be generated within the two Higgs doublet model.
Thus we will now investigate whether both baryogenesis and the explanation of the direct detection data via interference can be achieved simultaneously using a two Higgs doublet model. 

We start by introducing a DM-Higgs effective Lagrangian, which avoids the large DM-Higgs couplings discussed in the end of the last section.
The Lagrangian reads
 \begin{align}\label{DMlagr}
\mathscr{L}_{\rm DM} & = (m_{DM} - \frac{ \lambda^{DM}_1 v^2_1 }{ \Lambda } - \frac{ \lambda^{DM}_2 v^2_2 }{ \Lambda } ) \bar{\psi} \psi + \frac{ \lambda^{DM}_1 }{ \Lambda }  \phi^{\dagger}_1 \phi_1\bar{\psi} \psi + \frac{ \lambda^{DM}_2  }{ \Lambda } \phi^{\dagger}_2 \phi_2 \bar{\psi} \psi \, ,
 \end{align}
where the cut-off $\Lambda$ is assumed to be of the order of 1-10 GeV. We will show that for such a range of $\Lambda$ the required couplings to DM will turn to be between 1-10.  Here we indicated the Higgs doublets before EW symmetry breaking by $\phi$. In principle it is not hard to construct a UV complete theory for such a generic effective Lagrangian.
We give one such a model in the Appendix \ref{JV}.  
As an underlying two Higgs model we will use the one studied in~\cite{Cohen:1991iu,Nelson:1991ab,Joyce:1994zn}. After implementing the DM part, 
the full two Higgs model Lagrangian is 
\begin{align}
\mathscr{L}_H = \sum^{2}_{i=1} | D_{\mu} \phi_i|^2 - V(\phi_1, \phi_2) +  \mathscr{L}_{\rm DM} +  \mathscr{L}_{\rm fermions} + \mathscr{L}_{\rm Yuk} + \mathscr{L}_{\rm gauge} \ ,
\end{align}
where the two Higgs doublets scalar potential is 
 \begin{align}\label{2Higgspotential}
V(\phi_1, \phi_2) &= \lambda_1(\phi^{\dagger}_1 \phi_1 - v^2_1)^2 + \lambda_2(\phi^{\dagger}_2 \phi_2 - v^2_2)^2 +
\lambda_3[(\phi^{\dagger}_1 \phi_1 - v^2_1)+(\phi^{\dagger}_2 \phi_2 - v^2_2)]^2   \nonumber \\ 
&+\lambda_4[(\phi^{\dagger}_1 \phi_1)(\phi^{\dagger}_2 \phi_2 )  - (\phi^{\dagger}_1 \phi_2)(\phi^{\dagger}_2 \phi_1)] +
\lambda_5[{\rm{Re}}(\phi^{\dagger}_1 \phi_2) - v_1 v_2 \cos{\xi} ]^2    \nonumber \\
&+\lambda_6[{\rm{Im}}(\phi^{\dagger}_1 \phi_2) - v_1 v_2 \sin{\xi} ]^2 \,,
 \end{align}
$\xi$ being the relative CP violating phase between the two Higgs fields, which cannot be entirely rotated away by field redefinitions \cite{Branco:1979pv}. $\mathscr{L}_{\rm fermions}$ and $\mathscr{L}_{\rm gauge}$  account for the fermion covariant derivative terms and the gauge field kinetic terms respectively. Yukawa interactions in $ \mathscr{L}_{\rm Yuk}$ couple the up-type quarks to $ \phi_1$ and the down-type quarks to $ \phi_2$, resulting in identical Higgs-proton and Higgs-neutron couplings as presented in Eqs.~\eqref{twohiggs} and~\eqref{couplings1}.  The only relevant SM Yukawa coupling for baryogenesis is the top quark one \cite{Cohen:1991iu}. Due to the specific choice of the interaction between DM and the Higgs fields, baryogenesis is not affected by the presence of the DM sector. 
Fitting now the DM direct detection data using the model \eqref{DMlagr}, we get the following constraints
 \begin{align}\label{2Hconst1}
   &\tilde\lambda^{DM}_1 \left(\frac{\Lambda}{v_{EW}} \right)^{-1}
   \left(   \cos\theta - 4.0\, \sin\theta \frac{v_1}{v_2}\right)
    \left(\frac{m_{h_1}}{100\, \rm{GeV}} \right)^{-2} + \nonumber \\
   & 4.0 \, \tilde\lambda^{DM}_2   \left(\frac{\Lambda}{v_{EW}} \right)^{-1}  
    \left(   \cos\theta + 0.25\, \sin\theta \frac{v_2}{v_1}\right)
     \left(\frac{m_{h_2}}{100\, \rm{GeV}} \right)^{-2}
   = \pm 1.8 \times 10^2 \ ,
   \\
   \rule{0cm}{1cm}
   \label{2Hconst2}
  &\tilde\lambda^{DM}_1 \left(\frac{\Lambda}{v_{EW}} \right)^{-1}
  \left(   \cos\theta - 4.3 \sin\theta \frac{v_1}{v_2}\right)
    \left(\frac{m_{h_1}}{100\, \rm{GeV}} \right)^{-2}  + \nonumber \\
  & 4.3 \, \tilde\lambda^{DM}_2   \left(\frac{\Lambda}{v_{EW}} \right)^{-1}
    \left(   \cos\theta + 0.23 \sin\theta \frac{v_2}{v_1}\right)
     \left(\frac{m_{h_2}}{100\, \rm{GeV}} \right)^{-2}
   = \mp 1.3 \times 10^2  \ .
\end{align}
$\tilde\lambda^{DM}_1$ and $\tilde\lambda^{DM}_2$ are here defined so that $2 \tilde\lambda^{DM}_1 v_1 / \Lambda$ and $2 \tilde\lambda^{DM}_2 v_2 / \Lambda$ are the actual couplings of the DM to the physical Higgs fields $h_1$ and $h_2$, respectively,
\begin{equation}
\tilde\lambda^{DM}_1 = \lambda^{DM}_1 (\cos\theta - \sin\theta \frac{\lambda^{DM}_2}{\lambda^{DM}_1} \frac{v_2}{v_1}) \ ,
\qquad\qquad
\tilde\lambda^{DM}_2 = \lambda^{DM}_2 (\cos\theta +	\sin\theta \frac{\lambda^{DM}_1}{\lambda^{DM}_2} \frac{v_1}{v_2}) \ .
\end{equation}
Given the potential in Eq.~\eqref{2Higgspotential}, the mixing angle $\theta$ is now given by
\begin{equation}
\tan 2 \theta = \frac{2 v_1 v_2 ( 4 \lambda_3 + g)}{4 v^2_2 (\lambda_2 + \lambda_3) -4 v^2_1 (\lambda_1 + \lambda_3)+ g ( v^2_1- v^2_2)}  \ ,
\end{equation}
where $g = \lambda_5 \cos^2 \xi+ \lambda_6 \sin^2 \xi$.
If we assume that $\Lambda \sim 1$ GeV, and that both Higgs fields are light, $\mathscr{O}(100)$ GeV, we find that the DM couplings to the Higgs doublets $\lambda^{DM}_1$ and $\lambda^{DM}_2 $ (or at least one of them) need to be of the order $ \mathscr{O}(10)$ to be able to fulfill the above constraint equations.  The size of these couplings is now substantially reduced with respect to the previous model.   

Summarizing, since the CP violating phase can be rotated away in the light quark sector \cite{Branco:1979pv} there are no direct implications for the direct detection experiments. A welcome feature is that by including the interaction of the Higgs fields to DM using higher order operators, the energy scale $\Lambda$ can be traded for  a more natural value of the dimensionless couplings when fitting their values to direct detection data.

\section{Conclusions}

We have investigated several quantum mechanical interfering patterns for DM scattering off nuclei that can explain the DAMA and CoGeNT results. In particular we considered the case in which DM interacts via $Z$ and $Z'$, $Z'$ and  Higgs, and two Higgs fields with or without CP violation. We found that in the first case due to the constraints from the invisible decay width of the $Z$, the dominant contribution should come from the $Z'$ exchange. In the second case,  $Z'$ dominates again upon assuming natural values of the Higgs coupling and masses. In the last case we found that an explanation of the DAMA/CoGeNT results based on interference of two Higgs fields besides being phenomenologically viable,  is also consistent with the Electroweak Baryogenesis scenarios based on two Higgs doublet models. 

\appendix
\section{An ultraviolet completion  }
\label{JV}

In the Lagrangian below we introduce a simple renormalizable ultraviolet complete model for the effective theory presented in \eqref{DMlagr}: 
 \begin{align}  \label{Slagr}
\mathscr{L}_{S} & = \frac{1}{2} (\partial_{\mu} S)(\partial^{\mu} S) - \lambda_s(S^2-v_s^2)^2 +  y_{DM} S \bar{\psi} \psi  \nonumber  \\
                             &- y_1 [ (\phi^{\dagger}_1 \phi_1 - v^2_1) + (S^2-v_s^2) ]^2 - y_2 [(\phi^{\dagger}_2 \phi_2 - v^2_2)+ (S^2-v_s^2)]^2 \ , 
\end{align}
where $S$ is a new real, EW singlet, scalar field and $y_{DM}$, $y_1$, $y_2$ and $ \lambda_s$ are the dimensionless 
scalar-DM, scalar-Higgs 1, scalar-Higgs 2 and scalar self-couplings respectively.
The scalar potential, including the terms with couplings  $\lambda_s$,  $y_1$ and $y_2$, is minimized together with the two Higgs potential~\eqref{2Higgspotential}.  We define with $s$  the physical fluctuation of $S$ around its vev $v_s$.  
As long as the $\phi_i-s$ mixings in the Higgs-scalar mass matrix are tiny, the singlet field $s$ will couple mostly to the DM whereas the SM particles will couple mostly to the two Higgs fields. 

After integrating out the massive degrees of freedom, i.e. the Higgses and the scalar $S$, from the full model, the low energy effective theory describes also the four fermion interactions between the DM and the quarks, with effective couplings 
 \begin{align}
 & \frac{8 y_\text{DM} y_1 y_{q,1}  v_1 v_s } {m^2_s m^2_{h_1}} (\cos \theta  - \sin \theta \frac{y_2 v_2}{y_1 v_1}) (\cos \theta  - \sin \theta  \frac{y_{q,2}}{y_{q,1}} ) \ + \nonumber \\   
 & \frac{8 y_\text{DM} y_2 y_{q,2}  v_2 v_s } {m^2_s m^2_{h_2}} (\cos \theta  + \sin \theta \frac{y_1 v_1}{y_2 v_2}) (\cos \theta + \sin \theta  \frac{y_{q,1}}{y_{q,2}} ) \ ,
\end{align}
where $y_{q,i}$ is the quark-Higgs $i$ Yukawa coupling, before the diagonalization of the Higgs system, and $m_s$ and $m_{h_i}$ are the physical masses of  the singlet and the Higgs fields respectively. Matching the Lagrangian~\eqref{Slagr} with~\eqref{DMlagr} implies $4 y_\text{DM} y_i v_s / m_s^2 = \lambda_i^\text{DM} / \Lambda$. As long as the fundamental energy scale for $S$ is less than EW, i.e. $m_s \sim \Lambda \le v_{EW}$, the effective four fermion couplings can be much larger than the underlying couplings $y_{DM} $, $ y_i$ taken to be of the order of unity.

\end{document}